\documentclass[3p,times]{elsarticle}
\usepackage{ecrc}
\usepackage{amssymb}
\usepackage{microtype}
\usepackage{url}

\volume{00}
\firstpage{1}
\journalname{Nuclear Physics A}
\runauth{}
\jid{npa}
\jnltitlelogo{Nuclear Physics A}
\biboptions{square,comma,numbers,sort&compress}

\begin{document}

\begin{frontmatter}

\dochead{}

\title{Constraining nPDFs with inclusive pions and direct photons at forward rapidities in p+Pb collisions at the LHC}


\author[jyu,hip]{I.~Helenius}
\ead{ilkka.helenius@jyu.fi}
\author[jyu,hip]{K.~J.~Eskola}
\author[jyu,hip]{H.~Paukkunen}
\address[jyu]{Department of Physics, University of Jyv\"askyl\"a, P.O. Box 35, FI-40014 University of Jyv\"askyl\"a, Finland}
\address[hip]{Helsinki Institute of Physics, P.O. Box 64, FI-00014 University of Helsinki, Finland}

\begin{abstract}
In this talk, we present NLO pQCD predictions for inclusive pion and direct photon nuclear modifications in p+Pb collisions at mid- and forward rapidities at the LHC. In addition to the minimum bias predictions, we also address the centrality dependence with spatially dependent nuclear PDFs. To understand which regions of the nuclear momentum fraction $x_2$ these observables predominantly probe, we present also the underlying $x_2$ distributions at different rapidities. We are led to conclude that the isolated photons at forward rapidities are more sensitive to the small-$x_2$ dynamics than the inclusive pions.
\end{abstract}

\begin{keyword}
nuclear PDFs \sep centrality dependence \sep hard process \sep small $x$
\end{keyword}

\end{frontmatter}

\section{Introduction}
In the collinear factorization framework \cite{Collins:1989gx,Brock:1993sz} the cross section to produce a hard parton $k$ in a p+Pb collision can be calculated as
\begin{equation}
\mathrm{d} \sigma^{{\rm p} + {\rm Pb} \rightarrow k + X} = \sum\limits_{i,j,X'} f_{i}^{\rm p}(x_1,Q^2) \otimes f_{j}^{\rm Pb}(x_2,Q^2) \otimes \mathrm{d}\hat{\sigma}^{ij\rightarrow k + X'} + \mathcal{O}(1/Q^2),
\end{equation}
where the $f_{i}^{\rm p}(x_1,Q^2)$ $(f_{j}^{\rm Pb}(x_2,Q^2))$ are the parton distribution functions (PDFs) of the proton (lead nucleus) and $\mathrm{d}\hat{\sigma}^{ij\rightarrow k + X'}$ can be calculated at a given order of perturbative QCD (pQCD). This is valid when $Q^2\gg \Lambda_{\rm QCD}^2$ so that the higher-twist corrections are small. The PDFs are assumed to be universal and are usually extracted from experimental data in global analyses. While there are plenty of data to constrain the free proton PDFs \cite{Forte:2013wc}, the nuclear PDFs (nPDFs) are presently not well constrained and especially the uncertainties of the gluons are large at small $x$ \cite{Paukkunen:2014nqa}. The hope is that the forthcoming LHC data from p+Pb collisions could be used to bring the initial state nuclear modifications under better control and thereby also clarify their role in Pb+Pb measurements.

The global nPDF fits have so far considered only minimum bias nuclear collisions and it has not been possible to consistently use them in studies concerning the systematics of the nuclear collisions in different centrality classes. In Ref.~\cite{Helenius:2012wd} we developed a framework for the spatial dependence of the nPDFs, building on the $A$-dependence of the minimum bias nPDFs. A spatially dependent version of the EPS09 global fit \cite{Eskola:2009uj}, EPS09s, was also published. However, it has recently turned out that the experimental centrality determinations in p+$A$ collisions appear to impose significant, unwanted biases to the measured hard-process cross sections, the centrality effects being way larger than could have been envisaged from a sensible spatial dependence of the nPDFs. Our calculations here should therefore be seen as a baseline to such experimental findings.

With this short introduction, we now present our NLO pQCD predictions for the inclusive $\pi^0$ and direct photon nuclear modification ratios, discussing exactly which $x_2$ values are probed at different rapidities. The aim is to find whether one of these observables has a preference over the other in a sense that it would be more sensitive to small-$x_2$ gluons with more prominent nPDF-based centrality dependence.

\vspace{-0.3cm}
\section{Inclusive pion production}

\begin{figure}[htb]
\vspace{-0.2cm}
\begin{minipage}[t]{0.48\linewidth}
\centering
\includegraphics[width=\textwidth]{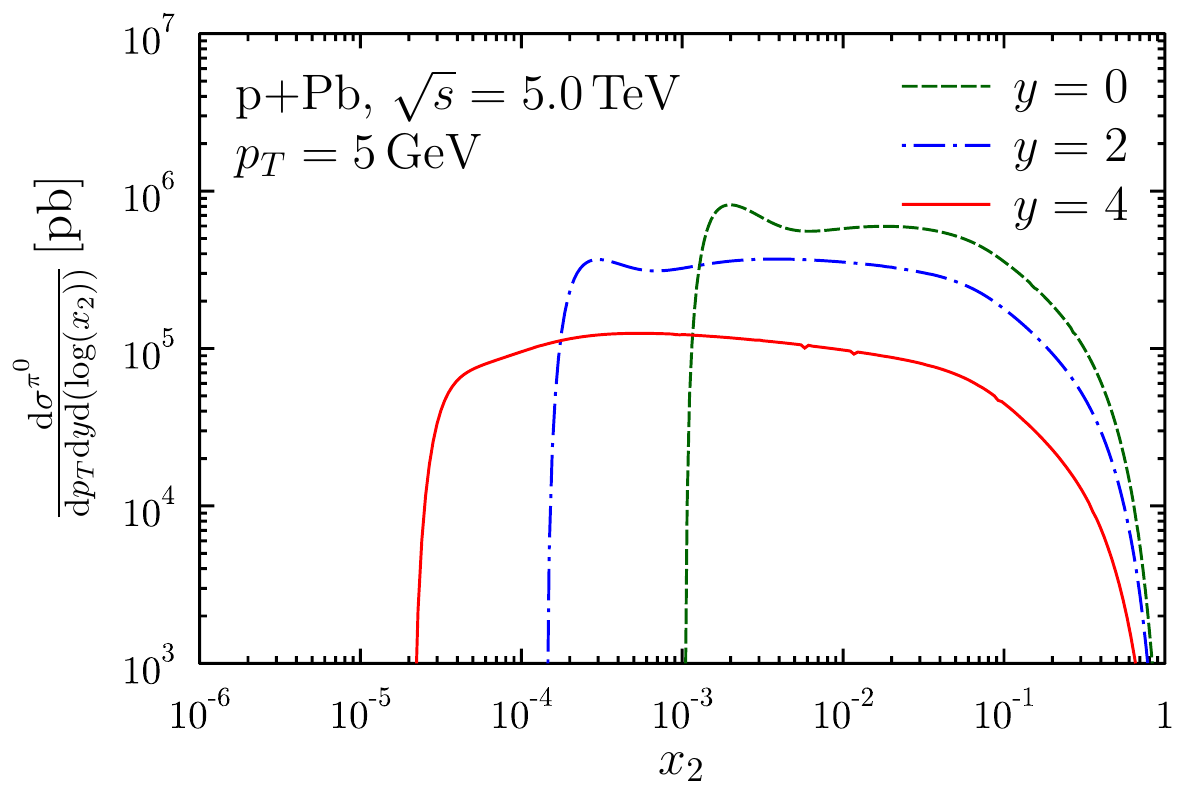}
\caption{The $x_2$ distribution for $\pi^0$ production in p+Pb collisions with $\sqrt{s_{NN}}=5.0\,\mathrm{TeV}$ at $p_T=5\,\mathrm{GeV/c}$ and $y=0$ (green dashed), $y=2$ (blue dot-dashed), and $y=4$ (red solid).}
\label{fig:dsigma_x2_pion}
\end{minipage}
\hspace{0.02\linewidth}
\begin{minipage}[t]{0.48\linewidth}
\centering
\includegraphics[width=\textwidth]{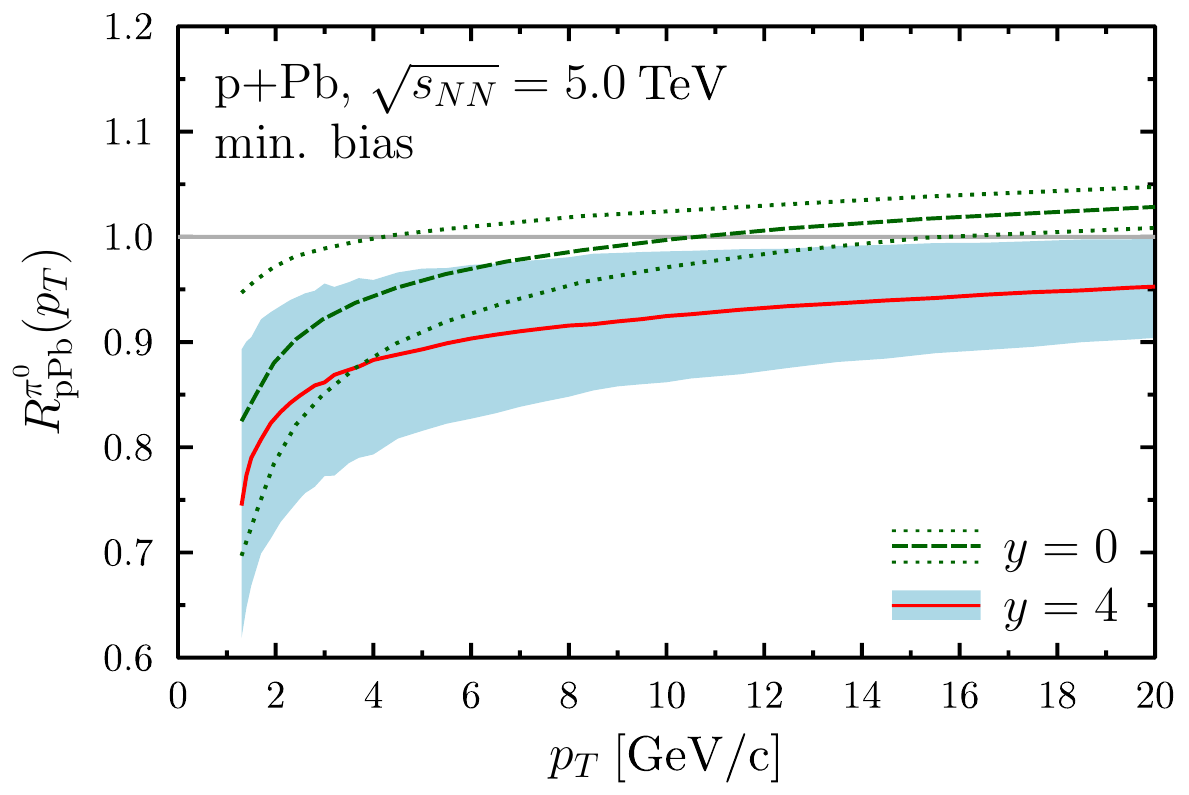}
\caption{The nuclear modification ratio $R_{\rm pPb}^{\pi^0}$ at $y=0$ (green dashed) and $y=4$ (red solid) using EPS09 NLO nPDFs. The blue uncertainty band for $y=4$ and the green dotted lines for $y=0$ are calculated from the EPS09 error sets.}
\label{fig:R_pPb_pi0_y45}
\end{minipage}
\end{figure}

The inclusive hadronic cross sections are calculated by convoluting the hard-parton spectra with the non-perturbative fragmentation functions (FFs). For the neutral pions this can be written as
\vspace{-3pt}
\begin{equation}
\mathrm{d} \sigma^{\pi^0+X}_{\rm pPb} = \sum_k \mathrm{d} \sigma^{k+X}_{\rm pPb} \otimes D_{\pi^0/k}(z,Q_F^2),
\label{eq:dsigma_h}
\vspace{-4pt}
\end{equation}
where $D_{\pi^0/k}(z,Q_F^2)$ is the parton-to-pion FF and $z$ is the fraction of the parent parton momentum carried by the pion. Due to this convolution, the transverse momentum of the hadron ($p_T$) does not have direct correspondence to the partonic kinematics and thus the cross section gets contributions from a wide range of $x_2$ even with fixed $p_T$ and rapidity $y$. The $x_2$ distributions for rapidities $y=0$, $y=2$, and $y=4$ with $\sqrt{s_{NN}}=5.0\,\mathrm{TeV}$ and $p_T=5.0\,\mathrm{GeV/c}$ are plotted in Figure~\ref{fig:dsigma_x2_pion}. The NLO calculations are done with \texttt{INCNLO}-code \cite{incnlopage,Aversa:1988vb,Aurenche:1987fs} using the CTEQ6.6M proton PDFs \cite{Nadolsky:2008zw} with the EPS09 nuclear modifications \cite{Eskola:2009uj} and DSS FFs \cite{deFlorian:2007aj}. The figure shows that even at $y=4$ there is a large contribution from $x_2>0.01$ which corresponds to the antishadowing region in EPS09. 
To see how the differences in the $x_2$ distributions at different rapidities map to the p+Pb $\pi^0$-production, the nuclear modification ratio $R_{\rm pPb}^{\pi^0}$ is plotted in Figure~\ref{fig:R_pPb_pi0_y45} for $y=0$ (originally published in Ref.~\cite{Helenius:2012wd}) and $y=4$. As the pions at forward rapidities probe smaller $x_2$ than the midrapidity pions, the suppression is correspondingly stronger due to the increasingly stronger shadowing in EPS09 at small $x_2$. Also the uncertainties are larger at the forward direction which reflects the lack of constraints for the gluon nPDFs.

\vspace{-1.0cm}
\section{Direct photon production}
\begin{figure}[htb]
\vspace{-0.2cm}
\begin{minipage}[t]{0.48\linewidth}
\centering
\includegraphics[width=\textwidth]{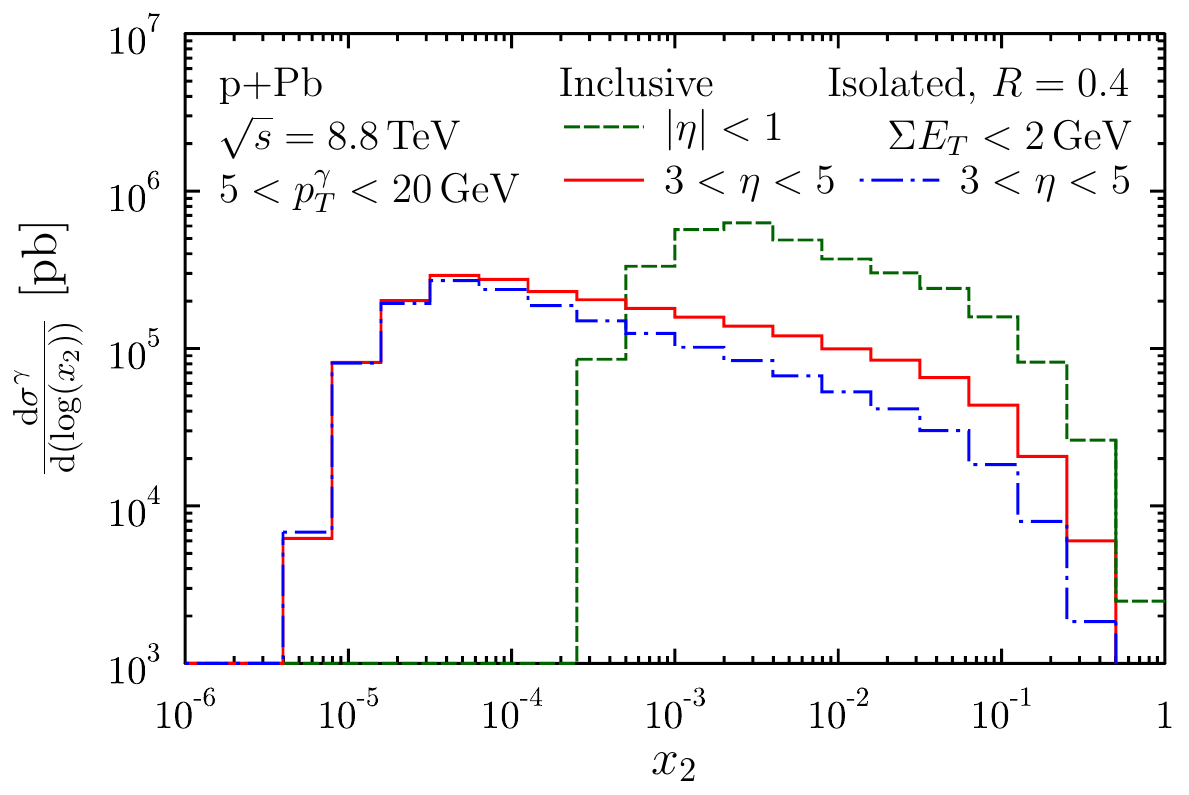}
\caption{The $x_2$ distribution for direct photons in p+Pb collisions with $\sqrt{s_{NN}}=8.8\,\mathrm{TeV}$ for $5<p_T<20\,\mathrm{GeV/c}$ at $|\eta|<1$ (green dashed) and $3<\eta<5$ (red solid), and for isolated photons at $3<\eta<5$ (blue dot-dashed).}
\label{fig:dsigma_x2_gamma}
\end{minipage}
\hspace{0.02\linewidth}
\begin{minipage}[t]{0.48\linewidth}
\centering
\includegraphics[width=\textwidth]{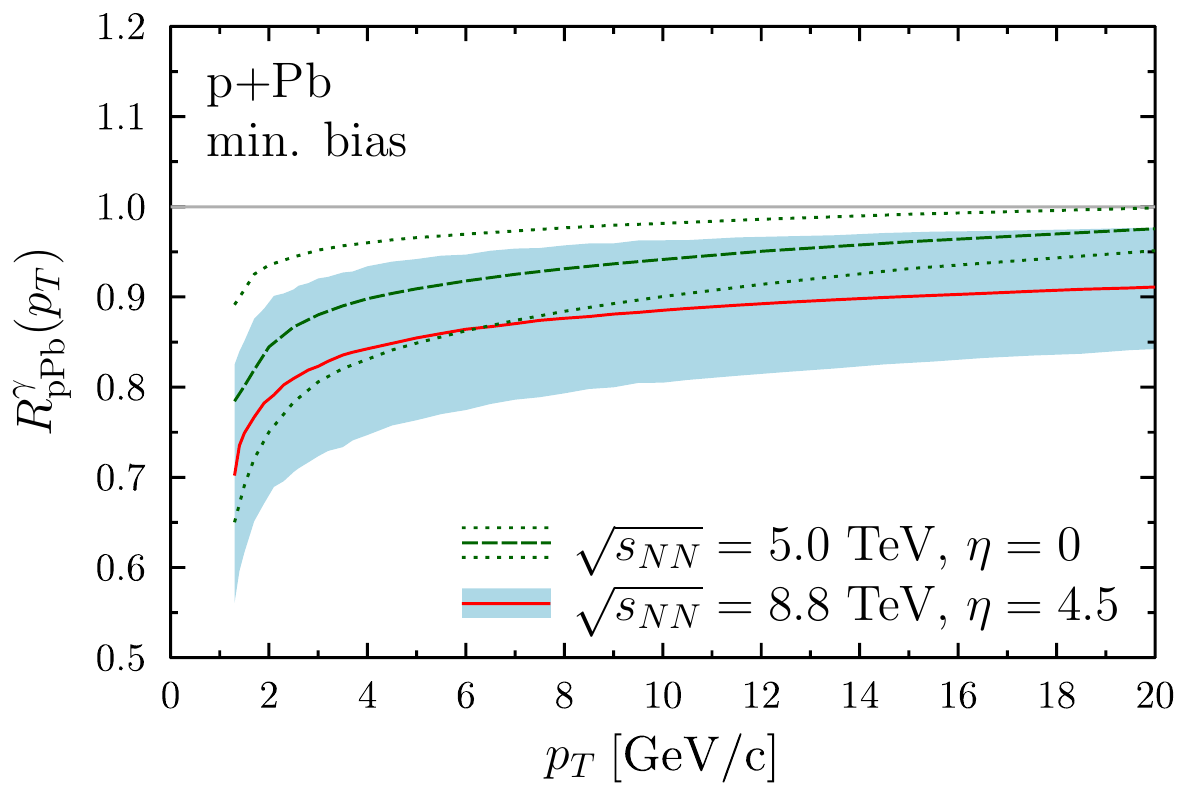}
\caption{The nuclear modification ratio $R_{\rm pPb}^{\gamma}$ for $\sqrt{s_{NN}}=5.0\,\mathrm{TeV}$ at $y=0$ (green dashed), published in Ref.~\cite{Helenius:2013bya}, and for $\sqrt{s_{NN}}=8.8\,\mathrm{TeV}$ at $y=4.5$ (red solid) using EPS09 NLO nPDFs. The blue uncertainty band for $\eta=4.5$ and the green dotted lines for $\eta=0$ are calculated from the EPS09 error sets.}
\label{fig:R_pPb_mb_gamma_y45}
\end{minipage}
\vspace{-0.1cm}
\end{figure}

The direct photons that can be measured in the experiments can originate from two different process: either they are produced directly in the hard partonic scattering (prompt photons) or they are produced similarly as high-$p_T$ hadrons by a fragmentation process. The prompt photons are more directly related to partonic kinematics and are thus more sensitive to small-$x_2$ values. However, the prompt and fragmentation components cannot be separated from each other in the measurements and thus both have to be included in the inclusive photon calculations. The net $x_2$ sensitivity of direct photons is plotted in Figure~\ref{fig:dsigma_x2_gamma} for p+Pb collisions with $\sqrt{s_{NN}}=8.8\,\mathrm{TeV}$ for $5 < p_T < 20\,\mathrm{GeV/c}$ at pseudorapidity ranges $|\eta|<1$ and $3<\eta<5$.
Although it has been shown \cite{d'Enterria:2012yj} that, at the LHC energies, the fragmentation component is dominant at low $p_T$, the presence of the prompt component makes the cross section somewhat more sensitive to smaller $x_2$ than the pions at corresponding rapidities above. The fragmentation component yields a similar tail towards higher $x_2$ values as for hadronic observables.
Even though the fragmentation component cannot be completely excluded, it can be suppressed by introducing an isolation cut for the direct photons. Here we have studied an isolation criterion which rejects photons that are accompanied by more than $2\,\mathrm{GeV}$ of hadronic energy inside a cone of radius $0.4$ around the photon. As can be seen from Figure~\ref{fig:dsigma_x2_gamma}, the isolation criterion indeed cuts off part of the tail at the high-$x_2$ region making the isolated photon cross section more sensitive to small $x_2$. The $x_2$ distributions are calculated here using the \texttt{JETPHOX} code \cite{jetphoxpage, Catani:2002ny, Aurenche:2006vj} and for the fragmentation component the BFGII parton-to-photon FFs \cite{Bourhis:1997yu} are used, otherwise the setup is the same as above.

Figure~\ref{fig:R_pPb_mb_gamma_y45} shows the nuclear modification factor for inclusive photons at mid- and forward rapidities. Despite the increased small-$x_2$ sensitivity, the suppression at forward rapidities is very similar to that of pions. This is mainly due to the rather mild $x_2$ dependence of the EPS09 nPDFs at small $x_2$, and for the rapid DGLAP evolution which tends to smooth out all $x_2$-dependent effects. At mid-rapidity the suppression is slightly stronger than the one for pions and persists up to higher $p_T$ values. Interestingly this is not only due to the smaller $x_2$ values probed by direct photons but also due to the fact that the direct photons are affected by the quark PDFs in which the shadowing is actually stronger than for gluons (and the antishadowing is missing) at the relevant $x$ values in EPS09. The isolated photons are not plotted here but the resulting $R_{\rm pPb}$ is very close to the inclusive photon $R_{\rm pPb}$ at forward rapidities due to the reasons mentioned above \cite{HEP2014}. The uncertainties are again larger at forward rapidities which calls for further constraints from measurements.

\begin{figure}[htbp]
\centering
\includegraphics[width=0.91\textwidth]{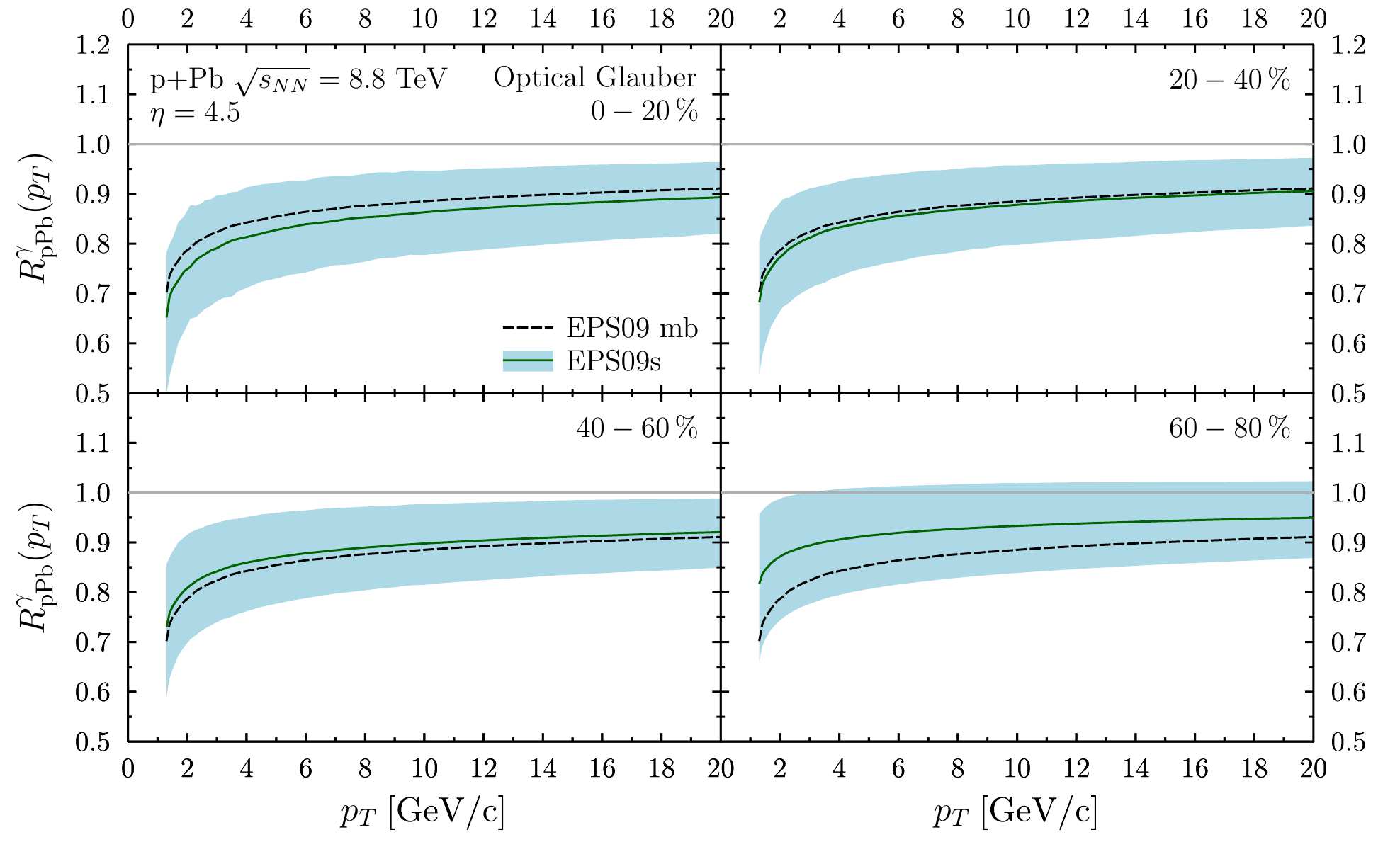}
\caption{The nuclear modification ratio $R_{\rm pPb}^{\gamma}$ for inclusive direct photon production in p+Pb collisions with $\sqrt{s_{NN}}=8.8\,\rm{TeV}$ at $\eta=4.5$ for different centrality classes (solid green) with EPS09s nPDFs and BFGII parton-to-photon FFs. The uncertainty bands are from EPS09s error sets. Also the minimum bias result (black dashed) from Figure \ref{fig:R_pPb_mb_gamma_y45} is plotted to each panel for comparison.}
\label{fig:R_pPb_gamma_y45}
\vspace{-0.3cm}
\end{figure}

The centrality dependence from EPS09s has already been observed to be rather mild at mid-rapidity \cite{Helenius:2012wd, Helenius:2013bya} but as the nuclear effects are stronger at forward rapidities, also the centrality dependence should be more pronounced there. To quantify this, the $R_{\rm pPb}$ for inclusive direct photon production is plotted in Figure~\ref{fig:R_pPb_gamma_y45} for centrality classes $0-20\,\%$, $20-40\,\%$, $40-60\,\%$, and $60-80\,\%$ at $\eta=4.5$. The centrality classes are defined in terms of impact parameters using the Optical Glauber model \cite{Helenius:2012wd}. The minimum bias result from Figure~\ref{fig:R_pPb_mb_gamma_y45} is plotted to each panel for comparison. Also here the centrality dependence seems to be rather mild --- the $10\,\%$ difference between the most central and the most peripheral could, perhaps, be eventually observed.

\vspace{-0.3cm}
\section{Conclusions}
Using NLO pQCD tools, we have shown that at forward rapidities the inclusive direct photons are more sensitive to small-$x_2$ physics than the inclusive hadrons at the same kinematical region, and that the sensitivity can be further improved by imposing an isolation cut for the direct photons. This suggests that the isolated photons at large rapidity would be a good observable to test the universality of the nPDFs at small $x_2$ and also to study their centrality dependence.
 
\vspace{-0.3cm}
\section*{Acknowledgments}
\noindent I.H. is supported by the Magnus Ehrnrooth foundation. The Academy of Finland, Project No. 133005, is acknowledged.

\bibliographystyle{elsarticle-num}
\bibliography{HP2013_proc}

\end{document}